\documentstyle[preprint,aps]{revtex}

\tighten
\input epsf

\begin{document}
\draft

\title{\Large \bf Dynamic Simulations of 
the Kosterlitz-Thouless Phase Transition}

\author{\bf B. Zheng$^{*\dagger}$, M. Schulz$^*$ and 
S. Trimper$^*$}

\address{$^*$ Universit\"at -- Halle, 06099 Halle, 
Germany}

\address{$^\dagger$ Universit\"at -- GH Siegen, 57068 Siegen, 
Germany}

\maketitle

\begin{abstract}
Based on the short-time dynamic scaling form,
a novel dynamic approach is proposed to tackle numerically
the Kosterlitz-Thouless phase transition.
Taking the two-dimensional XY model as an example,
the exponential divergence of the spatial correlation length,
the transition temperature $T_{KT}$
and all critical exponents are computed. 
Compared with Monte Carlo simulations in equilibrium,
we obtain data at temperatures nearer to $T_{KT}$.
\end{abstract}

\pacs{PACS: 02.70.Lq, 75.10.Hk, 64.60.Fr, 64.60.Ht}

The Kosterlitz-Thouless (KT) phase transition is 
an important kind of phase transitions in nature
\cite {kos73,kos74}.
When the temperature approaches the transition temperature
$T_{KT}$ from above,
the spatial correlation length diverges {\it exponentially},
rather than by a power law in a second order phase transition.
{\it Below $T_{KT}$, the system remains critical}
in the sense that the spatial correlation length is divergent.
No real long range order emerges in the whole temperature
regime. Important examples of systems with
a KT transition are the classical XY-type models, quantum
Heisenberg models, hard disk models and other 
relevant fluid systems as well as field theories.

It is well known that due to the exponential divergence
at the transition temperature, numerical simulations of
 critical systems with a KT transition suffer severely
from critical slowing down. 
For example, to compute the spatial correlation length
of the two-dimensional 
classical XY model, even with  
the cluster algorithm and the over-relaxed algorithm
one has only reached the temperature $T=0.98$,
which is still fairly far from 
$T_{KT}$ estimated to be around $0.89$ to $0.90$ 
\cite {gup88,wol89a,gup92}.
If some quenched randomness is added to the system,
e.g. in the fully frustrated XY model, 
the situation becomes even more complicated
\cite {gra93,lee94,ols95,jos96,ols96}.

Recently much progress has been made in critical
dynamics. 
It is discovered that universal dynamic scaling behavior
emerges already in the {\it macroscopic} short-time regime,
after a microscopic time scale $t_{mic}$ 
\cite {jan89,hus89,sta92,li94,gra95,sch95,maj96a,kre97,zhe98}.
More interesting and important is that the static
exponents originally defined
in equilibrium enter the short-time dynamic scaling.
This provides a possible way for extracting these exponents
from the short-time dynamic scaling behavior
\cite {zhe98}.
Since the measurements are carried out in
the macroscopic short-time regime,
the method is free of critical slowing down.

Such a short-time dynamic approach has been systematically
investigated for critical dynamic systems with {\it a second order
phase transition}. It has been first verified in the simple Ising
and Potts model \cite {li96,sch96}and  
recently applied successfully to 
 general and complex systems as non-equilibrium dynamic systems
 \cite {men98,mar98},
 the chiral degree of freedom
 in the fully frustrated XY model \cite {luo98} and 
 lattice gauge theory \cite {oka98}.
The critical exponents as well as
the critical temperature can be extracted either
from the power law behavior of the observables
at the early times or from the finite size scaling.
Compared with the non-local cluster algorithms,
the dynamic approach does study the dynamic properties 
of the original local dynamics.

However, it is not clear whether 
the short-time dynamic approach can systematically go beyond
critical systems with a second order phase transition,
even though first step approach or attempt has been made 
to the KT transitions and the spin glass
transitions 
\cite {luo97a,ber96,ber97,blu92}.
For systems with a KT
 transition, for example, owing to the {\it absence of symmetry breaking}
 and to the fact that
 the system remains critical below $T_{KT}$,
 a clear signal as for a second order phase transition
\cite {sch96,zhe98,luo98} does not exists
for the transition temperature $T_{KT}$.
The exponent $\nu$ and $T_{KT}$
have not been determined.
Standard techniques developed 
for the second order phase transitions do not apply here.
On the other hand, besides the exponents, whether
one can obtain other critical properties
of the equilibrium state as especially the spatial correlation length
from the short-time dynamics remains unknown,
{ \it even for second order phase transitions}.

In this communication we propose  
a short-time dynamic approach to the KT
 transition taking
the two-dimensional XY model as an example.
From the short-time dynamic scaling, we extract 
 the {\it spatial correlation length
of the  equilibrium state}. From the spatial correlation length
we estimate 
the transition temperature $T_{KT}$ and the static exponent
$\nu$. With $T_{KT}$ at hand, the static exponent $\eta$
and dynamic exponent $z$
are obtained from power law behavior of the magnetization
and Binder cumulant.
 There are no principle reasons
to choose the XY model but only because there exist
the most data of Monte Carlo simulations
in equilibrium for comparison.

The XY model in two dimensions
is defined by the Hamiltonian
\begin{equation}
H= \frac{1}{T}  \sum_{<ij>} \ \vec S_i \cdot \vec S_j\ ,
\label{e10}
\end{equation}
where $\vec S_i = (S_{i,x},S_{i,y})$ is a planar unit vector at site
$i$ and the sum is over the nearest neighbors.
In our notation, the coupling constant is already absorbed in the temperature.
Large scale Monte Carlo simulations in equilibrium
 have been performed to understand
the properties of the phase transition \cite {gup88,wol89a,gup92}.
The spatial correlation length $\xi$ and susceptibility
$\chi$ have been calculated
in a temperature interval $[0.98,1.43]$ 
with lattice sizes up to $512$.
The results support a KT singularity
for the spatial correlation length
\begin{equation}
\xi (\tau) \sim  exp\ (b \ \tau ^{-\nu})
\label{e20}
\end{equation}
and for the susceptibility
$\chi (\tau) \sim \xi ^{2-\eta} (\tau)$,
with $\tau \sim (T-T_{KT})/T_{KT}$ being the reduced temperature.
 However, unconstrained four-parameter fits
to the data do not yield 
completely satisfactory results \cite {gup92} .
The measured values of
$\nu$ and $T_{KT}$
from the data of $\xi$ and $\chi$ are
not very consistent and stable.
$\eta$ estimated from $\chi (\tau) \sim \xi ^{2-\eta} (\tau)$
 is above $0.7$ and 
too big compared with the theoretical prediction 
$\eta = 0.25$. 
The temperatures for the available data
of $\xi$ and $\chi$ are still far from the transition
temperature $T_{KT}$ estimated to be around
$0.89$ to $0.90$ 
(for details, see Table~\ref {t1} and Ref.~\cite {gup92}).
 However, simulations in equilibrium 
with lower temperatures are very difficult.

We will demonstrate that from the short-time dynamic scaling, 
the spatial correlation length
$\xi (\tau)$ of the  equilibrium state
 can be extracted
with relatively small lattices.
This is because the {\it non-equilibrium}
spatial length $\xi(t,\tau)$ is small
in the short-time regime of the dynamic evolution.
Therefore, simulations can be performed at lower temperatures.

In this paper we consider only the dynamics of model A,
which is relaxational without energy and magnetization
conservation. Starting from an {\it ordered initial
state}, e.g. all $\vec S_i = (S_{i,x},S_{i,y}) = (1,0)$,
the system is updated at the temperature above
$T_{KT}$ with the {\it standard Metropolis algorithm}.
We stop updating at a certain Monte Carlo time $t_m$
and repeat the procedure. Total samples for average
is from $800$ to $1\ 200$ for lattice size $L=256$ and 
above $400$ for $L=512$.
The lattice size $L=256$ is used in simulations
for temperatures from $T=1.07$ down to $0.975$,
while $L=512$ from $T=0.97$ to $0.94$.
Extra simulations with other lattice sizes confirm
that our data have no visible finite size effect.

The observable we measure is the magnetization
defined as 
\begin{equation}
M(t) = \frac{1}{L^d} <\ \sum_{i} \ S_{i,x} (t) \ > .
\label{e30}
\end{equation}
From a general physical view point of the renormalization 
group transformation, the magnetization $M(t)$ is subject
to a scaling form
\begin{equation}
M(t,\xi(\tau)) =t^{-\eta/2 z}M(1,t^{-1/z}\xi(\tau)).
\label{e40}
\end{equation}
When the temperature is at $T_{KT}$ (or below), i.e. $\tau=0$,
$\xi(\tau) \rightarrow \infty$ and $M(t)$ undergoes a power law
decay $M(t) \sim t^{-\eta/2 z}$.
However, for $\tau > 0$, the power law behavior is modified
by the scaling function $M(1,t^{-1/z}\xi(\tau))$.
This fact can be used for the determination $\xi^z (\tau)$
and the exponent $\eta/z$.

\begin{figure}[p]\centering
\epsfysize=6.cm
\epsfclipoff
\fboxsep=0pt
\setlength{\unitlength}{0.6cm}
\begin{picture}(9,9)(0,0)
\put(-2,-0.5){{\epsffile{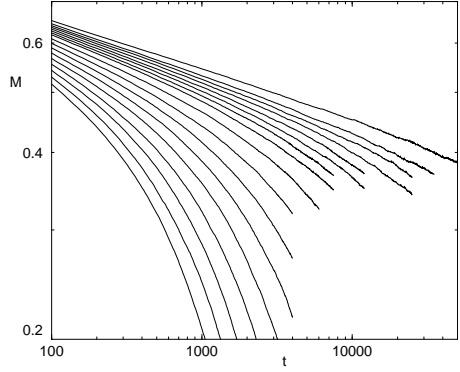}}}
\end{picture}
\caption{Time evolution of the magnetization in log-log scale. 
The temperatures are $0.94$, $0.95$, $0.955$, $0.96$, $0.965$, $0.97$,
 $0.975$, $0.98$, $0.99$, $1.00$, $1.01$, $1.02$, $1.03$,
  $1.04$, $1.05$, $1.06$, $1.07$      
 (from above).}
\label{f1}
\end{figure}

In Fig.~\ref {f1}, the time evolution of the magnetization
is displayed in log-log scale for different temperatures.
We perform Monte Carlo simulations up to a time $t_m$
where there is visible deviation from the power law behavior.
Actually, in the short-time regime of the dynamic evolution,
the magnetization itself is more or less self-averaged.
We may increase the lattice size without too much
extra fluctuation. What restricts our simulations
to very low temperatures is only $t_m$.
Our longest updating time
is $t_m =50\ 000$ at the temperature $T=0.94$.
To obtain the curve for $T=0.98$ one needs $8$ days
in ALPHAstation $500$ (400MHz), while ten times more
for $T=0.94$.

\begin{figure}[p]\centering
\epsfysize=6.cm
\epsfclipoff
\fboxsep=0pt
\setlength{\unitlength}{0.6cm}
\begin{picture}(9,9)(0,0)
\put(-2,-0.5){{\epsffile{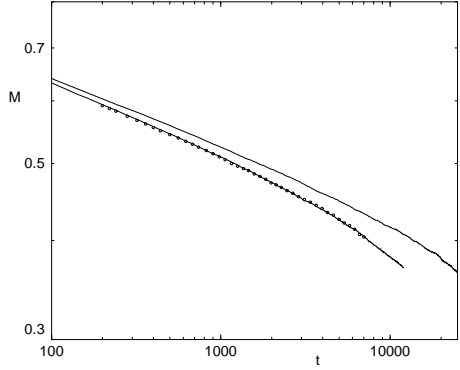}}}
\end{picture}
\caption{The scaling plot of the magnetization with a pair of temperatures
$(T_1,T_2)$. The upper and lower solid lines correspond to 
temperatures $T_1=0.955$ and $T_2=0.965$. The circles are
also the magnetization
with $T_1=0.955$ but rescaled to have the best fit with that of $T_2=0.965$.}
\label{f2}
\end{figure}

Now from a scaling collapse of two curves
with a pair of temperatures $(T_1,T_2)$, 
we estimate the ratio $\xi^z_1/\xi^z_2$ and $\eta/z$.
Here $\xi_1$ and $\xi_2$ are the values of $\xi (\tau)$
at the temperatures $T_1$ and $T_2$ respectively.
In Fig. \ref {f2}, such a scaling plot is displayed
for $(T_1,T_2)=(0.955,0.965)$.
We multiply the magnetization $M(t_1,\xi_1)$ by an overall factor
$b^\alpha$ and rescale $t_1$ to $t_1/b$. 
According to the scaling form (\ref {e40}) 
this rescaled $M(t_1,\xi_1)$ is equal
to $M(t_2,\xi_2)$ if and only if $b=\xi^z_1/\xi^z_2$ and $\alpha=\eta/2z$.
Therefore, searching for the best fit between $M(t_2,\xi_2)$
and the rescaled $M(t_1,\xi_1)$ we determine
$\xi^z_1/\xi^z_2$ and $\eta/2z$.
In the figure, the circles represents
the rescaled $M(t_1,\xi_1)$ best fitted to $M(t_2,\xi_2)$.

Theoretically the exponents $\eta$ and $z$ 
are defined at the transition temperature $T_{KT}$.
 The exponent $\nu$ and the exponential singularity
in Eq.~(\ref {e20})
are defined for temperatures above but in the close neighborhood
of $T_{KT}$. If the temperature is fairly above $T_{KT}$,
in principle, all the exponents and the parameters $b$
may have some dependence on the temperature.
Usually this dependence on the temperature is neglected,
 otherwise the situation becomes too complicated
 \cite {gup92}.
  In our dynamic approach, we perform the scaling
collapse of the magnetization with two temperatures
which are not too far away each other and therefore
the dependence of $\eta/z$ on the temperature are actually considered.
 However, we assume $\nu$ and $b z$ be independent of 
the temperature.

In Table~\ref {t2}, the measured ratio of $\xi^z_1/\xi^z_2$
and exponent $\eta/2z$ for
different pairs of temperatures $(T_1, T_2)$ are listed.
Errors are estimated by dividing the total samples 
for the time-dependent magnetization into two groups only.
The lowest temperature we reach is $T=0.94$.
For comparison, in Table~\ref {t2} available 
values of $\xi^z_1/\xi^z_2$
from Ref.~\cite {gup92} (denoted by $\dagger$)
are also given. They are slightly smaller than our results.
We observe that as the temperature decreases,
these values of ${\xi^z_1/\xi^z_2}^\dagger$ do not 
increase sufficiently smoothly.
Real errors of these data
(and also our data) might be somehow bigger
than given in the table.
Taking $z=1.96$ in Table~\ref {t1} as input, 
the resulting $\eta$ from $\eta/2z$ in Table~\ref {t2}
is around $0.26$ to $0.31$.
 As expected \cite {gup92}, we see a tendency that
 $\eta$ will become around $0.25$ 
 as the temperature approaches $T_{KT}$.

Now we fit the data in Table~\ref {t2} to the exponential form 
in Eq.~(\ref {e20}) and estimate the transition temperature $T_{KT}$,
the exponent $\nu$ and parameter $bz$.
The best results are given in 
Table~\ref {t1} in comparison with
the those from simulations in equilibrium.
Our results are fitted from a relatively lower
temperature interval $[0.94, 1.07]$,
and agree well with those obtained
in a temperature interval $[0.98, 1.43]$
in Ref.~\cite {gup92}, in both cases of a unconstrained fit and
a fit with a fixed $\nu=0.5$.
However, as pointed out by the authors of Refs.\cite {gup92,wol89a},
for the unconstrained fit the minimum in the parameter
space is not very stable in the directions
of $\nu$ and $bz$. It is somehow by chance that
our value of $\nu=0.48$ is so close to $\nu=0.47$
 obtained in Ref.~\cite {gup92}.
 When we vary $T_2$ in the fitting interval
 $[0.94, T_2]$ from $1.07$ to smaller values, 
 the exponent $\nu$ first drops down and 
 then rises again after around $T_2=1.00$.
 These fluctuations very probably come
 from the fact that we do not have sufficient data points
and accuracy for each data point.
The estimate of $T_{KT}$ is relatively stable.
But the impression is that the value of $T_{KT}$ 
might be slightly bigger than $T_{KT}=0.8942$
given in the table, if the fitting can confidently
be performed at the really close neighbourhood of
 $T_{KT}$.

\begin{figure}[p]\centering
\epsfysize=6.cm
\epsfclipoff
\fboxsep=0pt
\setlength{\unitlength}{0.6cm}
\begin{picture}(9,9)(0,0)
\put(-2,-0.5){{\epsffile{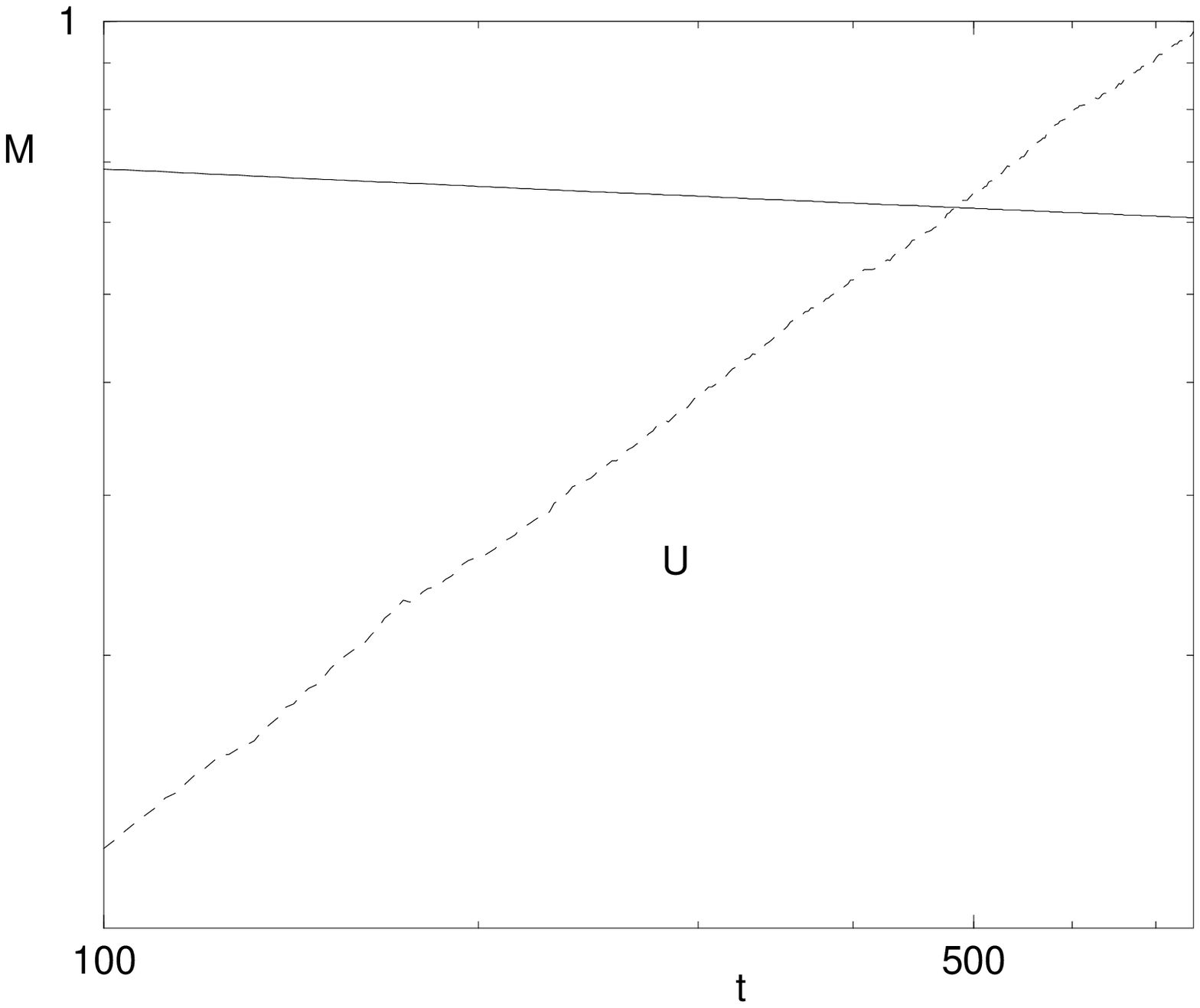}}}
\end{picture}
\caption{Time evolution of the magnetization $M$ and Binder cumulant $U$
at $T_{KT}=0.894$ in log-log scale. To plot the figure,
$U$ has been multiplied by a constant $200$.
} 
\label{f3}
\end{figure}

With the transition temperature $T_{KT}$ at hand,
we proceed to measure the magnetization $M$
and its second moment $M^{(2)}$ at $T_{KT}$. 
To seek for the dynamic exponent $z$, we construct
a Binder cumulant $U=M^{(2)}/M^2-1$.
Finite size scaling analysis leads to the short-time behavior
\cite {luo97a,zhe98,luo98}
\begin{equation}
U(t) \sim t^{d/z}.
\label{e50}
\end{equation}
Finally an accurate value of $\eta/2z$ can be obtained
from the power law decay of the magnetization
$M(t) \sim t^{-\eta/2z}$ (see Eq.~(\ref {e40})).
In Fig.~\ref {f3}, $M(t)$ and $U(t)$ have been plotted
in log-log scale. 
Since our measurements only extend to $t=750$,
a lattice size $L=64$ is sufficient. Total samples
for average is $12\ 000$.
From the slopes of the curves in the figure
we measure $d/z$ and $\eta/2z$, then calculate
$z$ and $\eta$. The results are included in Table~\ref {t1}.
Our $\eta=0.238(4)$ coincides with the best estimate $\eta=0.235(5)$
in equilibrium.

In conclusions, a dynamic approach is proposed to tackle numerically
 the Kosterlitz-Thouless phase transition.
 We demonstrate for the first time that not only the critical exponents
 but also the spatial correlation length of the  equilibrium state
 can be obtained from the short-time dynamics.
Taking the two-dimensional XY model as an example,
 the exponential divergence of the spatial correlation length
 is extracted from the short-time 
 dynamic scaling. The transition temperature $T_{KT}$,
 the static exponents $\nu$ and $\eta$ as well as
 the  dynamic exponent
 $z$  are then estimated. Since the measurements are carried out in the
 short-time regime
 of the dynamic evolution, where the {\it non-equilibrium} spatial correlation
 length is small, we do not encounter difficulties
 of generating independent configurations.
  Compared with the  
 simulations in equilibrium, we can perform simulations
 at the temperatures closer to $T_{KT}$.
   This method can in principle be applied or generalized to
 other kinds of phase transitions as second order phase transitions
 and spin glass transitions.

{\bf Acknowledgements}:
Work supported in part by the Deutsche Forschungsgemeinschaft;
Schu 95/9-1 and SFB~418.

\begin{table}[h]\centering
\begin{tabular}{ccccc}
 $[T_1, T_2]$  &$[0.94, 1.07]$ &$[0.94, 1.07]$   &$[0.98, 1.43]^\dagger$ &$[0.98, 1.43]^\dagger$  \\ 
\hline
 $T_{KT}$      & 0.8942& 0.8926  & 0.8953  &  0.8914  \\
 $b z$         & 4.12  & 3.82    & 3.67    &  3.38  \\
 $\nu$         & 0.48  & 0.5     & 0.47    &  0.5   \\
\hline
 $T^*_{KT}$    &       &         & 0.8871   &  0.8961  \\
 $\nu^*$       &       &         & 0.57    &  0.5  \\
\hline
 $z$           & 1.96(3) &       &         &  \\
 $\eta$      &   .238(4) &       & .235(5) &   \\
\end{tabular}
\caption{Our results of the exponents and $T_{KT}$
obtained in temperature interval $[T_1, T_2]$
in comparison with those in Ref.~\protect\cite {gup92}
(denoted by $\dagger$).
$T^*_{KT}$ and $\nu^*$ of Ref.~\protect\cite {gup92}
are from data of the susceptibility.
The second and fourth column are the results 
with a fixed $\nu=0.5$ as input.
Our values of
$z$ and $\eta$ are measured at $T_{KT}=0.894$
and $\eta$ of Ref.~\protect\cite {gup92} is estimated
 with finite size scaling and Monte Carlo
renormalization group methods at also
$T_{KT}=0.894$.
$bz$ for Ref.~\protect\cite {gup92} is calculated
by taking our $z$ as input.
}
\label{t1}
\end{table}

\begin{table}[h]\centering
\begin{tabular}{ccccccc}
    $(T_1, T_2)$  &(.940, .950)&(.950, 0.955)&(.950, 0.960)&(.955, .960)&(.955, .965)&(.960, .965) \\
$\xi^z_1/\xi^z_2$ & 5.75(38)   &  2.04(06)   & 3.61(07)    &   1.81(09) &  3.05(06)  &    1.65(04)  \\
    $\eta/2z$     & .0680(11)  &  .0695(16)  & .0690(17)   & .0682(21)  & .0674(12)  &  .0650(21)   \\
\hline
 $(T_1, T_2)$     &(.960, .970)&(.965, .970)&(.965, .975) &(.970, .975)&(0.97, 0.98)&(0.97, 0.99) \\
$\xi^z_1/\xi^z_2$ & 2.65(07)   &  1.56(03)& 2.35(02)   &  1.51(01)   & 2.165(45)  & 4.35(11)         \\
 $\eta/2z$        &0.0671(21)  & .0691(18) &   .0676(12) & .0685(08)& .0682(17)  & .0699(22)       \\
\hline
 $(T_1, T_2)$     &(0.975, 0.98)&(0.98, 0.99)&(0.98, 1.00)&(0.99, 1.00)&(0.99, 1.01)&(1.00, 1.01)\\
$\xi^z_1/\xi^z_2$ &  1.47(03)  & 1.965(43)   & 3.67(07)   &  1.840(40)   &   3.22(06) &  1.710(22)  \\
$\xi^z_1/\xi_2^{z\ \dagger}$ 
                  &            & 1.839(84)   &             &1.605(61)    &            &  1.600(39) \\
 $\eta/2z$        & .0668(20) & .0727(28)     &  .0768(15)  &.0753(22)   & .0774(14)  &  .0762(26)  \\
\hline
 $(T_1, T_2)$     &(1.00, 1.02)&(1.01, 1.02)&(1.01, 1.03)&(1.02, 1.03)&(1.02, 1.04)&(1.03, 1.04) \\
$\xi^z_1/\xi^z_2$ &  2.690(29)  & 1.564(08)  &  2.334(21)& 1.474(10)  & 2.131(20)  & 1.424(07)      \\
$\xi^z_1/\xi_2^{z\ \dagger}$ 
                  &             & 1.453(36)  &           &1.434(34)   &            &  1.351(38)\\
 $\eta/2z$        &  .0784(24)  & .0781(11)  &   .0768(19) & .0773(13)  & .0777(11)  & .0754(16) \\
\hline
 $(T_1, T_2)$     &(1.03, 1.05)&(1.04, 1.05)&(1.04, 1.06)&(1.05, 1.06)&(1.05, 1.07)&(1.06, 1.07) \\
$\xi^z_1/\xi^z_2$ &  1.964(17) &  1.380(07)& 1.832(19)  &  1.329(10)  &  1.726(13)  &  1.312(6)       \\
 $\eta/2z$        &  .0778(18) &.0835(42)   & .0777(12) &  .0705(46)&  .0781(19)  & .0900(24)      \\
\end{tabular}
\caption{
The measured ratio $\xi^z_1/\xi^z_2$ and exponent $\eta/2z$ for different pairs of
temperatures. Values of $\xi^z_1/\xi_2^{z\ \dagger}$ are calculated from
data in the table VIII of Ref.~\protect\cite {gup92}.
}
\label{t2}
\end{table}

\end{document}